\documentclass[twocolumn]{aastex63}

\usepackage{amsmath}

\newcommand{\kms}{{\rm km\,s}\ensuremath{^{-1}}}

\newcommand{\radioqso}{P172+18}

\newcommand{\lya}{\mathrm{Ly}\ensuremath{\alpha}}
\newcommand{\nv}{\ion{N}{5}}

\newcommand{\mgii}{\ion{Mg}{2}}
\newcommand{\civ}{\ion{C}{4}}
\newcommand{\siiv}{\ion{Si}{4}}

\newcommand{\ciii}{\ion{C}{3}]}
\newcommand{\feii}{\ion{Fe}{2}}

\newcommand{\oiv}{[\ion{O}{4}]}

\newcommand{\ips}{\ensuremath{i_{\rm P1}}}

\newcommand{\zps}{\ensuremath{z_{\rm P1}}}

\newcommand{\yps}{\ensuremath{y_{\rm P1}}}

\newcommand{\gde}{\ensuremath{g_{\rm DE}}}
\newcommand{\rde}{\ensuremath{r_{\rm DE}}}
\newcommand{\zde}{\ensuremath{z_{\rm DE}}}

\newcommand{\Jnot}{\ensuremath{J_{\rm NOT}}}
\newcommand{\Hnot}{\ensuremath{H_{\rm NOT}}}
\newcommand{\Knot}{\ensuremath{Ks_{\rm NOT}}}

\newcommand{\Jtmass}{\ensuremath{J_{\rm 2MASS}}}
\newcommand{\Htmass}{\ensuremath{H_{\rm 2MASS}}}
\newcommand{\Ktmass}{\ensuremath{K_{\rm 2MASS}}}

\accepted{January 31, 2021}

\shorttitle{A radio-loud quasar at $z=6.8$}
\shortauthors{Ba\~nados et al.}

\graphicspath{{./}{figures/}}

\begin{document}

\title{The discovery of a highly accreting, radio-loud quasar at $z=6.82$}

\correspondingauthor{Eduardo Ba\~nados}
\email{banados@mpia.de}

\author[0000-0002-2931-7824]{Eduardo Ba\~nados}
\affiliation{{Max Planck Institut f\"ur Astronomie, K\"onigstuhl 17, D-69117, Heidelberg, Germany}}
\affiliation{The Observatories of the Carnegie Institution for Science, 813 Santa Barbara Street, Pasadena, CA 91101, USA}
\author[0000-0002-5941-5214]{Chiara Mazzucchelli}\thanks{ESO Fellow}
\affiliation{European Southern Observatory, Alonso de Cordova 3107, Vitacura, Region Metropolitana, Chile}

\author[0000-0003-3168-5922]{Emmanuel Momjian}
\affiliation{National Radio Astronomy Observatory, Pete V. Domenici Science Operations Center, P.O. Box O, Socorro, NM 87801, USA}

\author[0000-0003-2895-6218]{Anna-Christina Eilers}\thanks{NASA Hubble Fellow}
\affiliation{MIT Kavli Institute for Astrophysics and Space Research, 77 Massachusetts Ave., Cambridge, MA 02139, USA}

\author[0000-0002-7633-431X]{Feige Wang}\thanks{NASA Hubble Fellow}
\affiliation{Steward Observatory, University of Arizona, 933 North Cherry Avenue, Tucson, AZ 85721, USA}

\author[0000-0002-4544-8242]{Jan-Torge Schindler}
\affiliation{{Max Planck Institut f\"ur Astronomie, K\"onigstuhl 17, D-69117, Heidelberg, Germany}}

\author[0000-0002-7898-7664]{Thomas Connor}
\affiliation{Jet Propulsion Laboratory, California Institute of Technology, 4800 Oak Grove Drive, Pasadena, CA 91109, USA}

\author[0000-0001-6102-9526]{Irham Taufik Andika}
\affiliation{{Max Planck Institut f\"ur Astronomie, K\"onigstuhl 17, D-69117, Heidelberg, Germany}}
\affiliation{International Max Planck Research School for Astronomy \& Cosmic Physics at the University of Heidelberg, Germany}

\author[0000-0002-3026-0562]{Aaron J. Barth}
\affiliation{Department of Physics and Astronomy, 4129 Frederick
  Reines Hall, University of California, Irvine, CA, 92697-4575, USA}
\author[0000-0001-6647-3861]{Chris Carilli}
\affiliation{National Radio Astronomy Observatory, Pete V. Domenici Science Operations Center, P.O. Box O, Socorro, NM 87801, USA}
\affiliation{Astrophysics Group, Cavendish Laboratory, J.J.\ Thomson Avenue, Cambridge CB3 0HE, UK}

\author[0000-0003-0821-3644]{Frederick B.\ Davies}
\affiliation{{Max Planck Institut f\"ur Astronomie, K\"onigstuhl 17, D-69117, Heidelberg, Germany}}

\author[0000-0002-2662-8803]{Roberto Decarli}
\affiliation{INAF --- Osservatorio di Astrofisica e Scienza dello Spazio, 
via Gobetti 93/3, I-40129, Bologna, Italy}

\author[0000-0001-5287-4242]{Xiaohui Fan}
\affil{Steward Observatory, University of Arizona, 933 N Cherry Ave, Tucson, AZ, USA}

\author[0000-0002-6822-2254]{Emanuele Paolo Farina}
\affiliation{Max Planck Institut f\"ur Astrophysik, Karl-Schwarzschild-Stra{\ss}e 1, D-85748, Garching bei M\"unchen, Germany}

\author[0000-0002-7054-4332]{Joseph~F.~Hennawi}
\affiliation{Department of Physics, Broida Hall, University of California, Santa Barbara, CA 93106--9530, USA}

\author[0000-0001-9815-4953]{Antonio Pensabene}
\affiliation{Dipartimento di Fisica e Astronomia, Alma Mater Studiorum, Università di Bologna, Via Gobetti 93/2, I-40129 Bologna, Italy.}
\affiliation{INAF --- Osservatorio di Astrofisica e Scienza dello Spazio, 
via Gobetti 93/3, I-40129, Bologna, Italy}

\author[0000-0003-2686-9241]{Daniel Stern}
\affiliation{Jet Propulsion Laboratory, California Institute of Technology, 4800 Oak Grove Drive, Pasadena, CA 91109, USA}

\author[0000-0001-9024-8322]{Bram P.\ Venemans}
\affiliation{{Max Planck Institut f\"ur Astronomie, K\"onigstuhl 17, D-69117, Heidelberg, Germany}}

\author[0000-0001-5245-2058]{Lukas Wenzl}
\affiliation{Department of Astronomy, Cornell University, Ithaca, NY 14853, USA}
\affiliation{{Max Planck Institut f\"ur Astronomie, K\"onigstuhl 17, D-69117, Heidelberg, Germany}}

\author[0000-0001-5287-4242]{Jinyi Yang}
\altaffiliation{Strittmatter Fellow}
\affil{Steward Observatory, University of Arizona, 933 N Cherry Ave, Tucson, AZ, USA}

\begin{abstract}
Radio sources at the highest redshifts can provide unique information on the first massive galaxies and black holes, the  densest
primordial environments, and the epoch of reionization. The number of astronomical objects identified at $z>6$ has increased 
dramatically over the last few years, but previously only three radio-loud
($R_{2500}=f_{\nu,5\,\mathrm{GHz}} / f_{\nu,2500\,\text{\AA}} >10$) sources had been reported
at $z>6$, with the most distant being a quasar at $z=6.18$. 
Here we present the discovery and characterization of PSO~J172.3556+18.7734, a radio-loud quasar at $z=6.823$. This source has an \mgii-based black hole mass of $\sim 3\times 10^8\,M_\odot$ and is one of the fastest accreting quasars, consistent with super-Eddington accretion. The ionized region around the quasar is among the largest measured at these redshifts, implying an active phase longer than the average lifetime of the $z\gtrsim 6$ quasar population. 
From archival  data, there is evidence that its 1.4\,GHz emission has decreased by a factor of two over the last two decades.  
The quasar's radio spectrum  between 1.4 and 3.0\,GHz is steep ($\alpha = -1.31$).  Assuming the measured radio slope and extrapolating to rest-frame 5\,GHz, the quasar has a radio-loudness parameter $R_{2500} \sim 90$. 
A second steep radio source ($\alpha=-0.83$) of comparable brightness to the quasar is only 23\farcs1 away ($\sim$120 kpc at $z=6.82$; projection probability $<2\%$), but shows no optical or near-infrared counterpart.  Further follow-up is required to establish whether these two sources are physically associated.

\end{abstract}


\section{Introduction} \label{sec:intro}

Radio jets from active galactic nuclei (AGNs) are thought to play a key role in the coevolution of supermassive black holes and their host galaxies, as well as in the early growth of massive black holes \citep[e.g.,][]{jolley2008, volonteri2015,hardcastle2020}.   Yet, strong radio emission seems to be a rare or at least short-lived phenomenon. Only about 10\% of all quasars are strong radio emitters, almost independent of their redshifts up to $z\sim 6$ (e.g., \citealt{banados2015a, yang2016, shen2019} but see also \citealt{jiang2007, kratzer2015}).

The radio-loudness of a quasar is usually defined as the ratio of rest-frame 5\,GHz (radio) and 4400\,\AA\ (optical) flux densities ($R_{4400}$; e.g., \citealt{kellermann1989}) although sometimes the  2500\,\AA\ (UV)  emission is used instead of the optical flux density ($R_{2500}$; e.g., \citealt{jiang2007}). For an unobscured type-1 quasar, the different definitions yield comparable results. 
An object is considered radio-loud\footnote{When we talk about radio-loud quasars in this paper we refer to jetted-quasars, see discussion in \cite{padovani2017}} if $R_{2500}$ or $R_{4400}$ is greater than 10. Radio-loud sources at the highest accessible redshifts are of particular interest for multiple reasons. For example, radio galaxies are known to be good tracers of overdense environments \citep[e.g.,][]{venemans2007b,wylezalek2013} and at high redshift these overdensities could be 
the progenitors of the galaxy clusters seen in the present-day universe \citep{overzier2016,noirot2018}. 
Furthermore, radio-loud sources deep in the epoch of reionization would enable crucial absorption studies of the intergalactic medium (IGM) at this critical epoch \citep[e.g.,][]{carilli2002,
ciardi2013,thyagarajan2020} and they could potentially also constrain the nature of dark matter particles by detecting  neutral hydrogen in absorption in the radio spectrum  \citep[e.g.,][]{shimabukuro2020}.

The number of astronomical objects known within the first billion years of the universe has increased dramatically over the last few years, with galaxies being discovered up
to $z\sim11$ \citep{oesch2014} and quasars up to $z\sim7.5$ \citep{banados2018a,yang2020}. On the other
hand, identifying strong radio emitters at high redshift has been difficult. The highest-redshift radio galaxy lies at $z=5.7$ \citep{saxena2018}, with the previous record at $z = 5.2$ \citep{vanbreugel1999}. 
Out of the 200 published quasars at $z>6$ \citep[e.g.,][]{banados2016, Matsuoka2019ApJ...883..183M,
andika2020}, only three are known to be radio-loud. For the large majority of the remainder, the existing radio data are too shallow to robustly classify them as radio-quiet or radio-loud, although there are on-going efforts to obtain deeper radio observations of these objects.  The three $z>6$ radio-loud quasars currently known\footnote{The quasar J1609+3041 at $z=6.14$ was classified as radio-loud by \cite{banados2015a}  based  on a tentative 1.4\,GHz detection at S/N of 3.5. However, deeper observations showed this object to be radio-quiet \citep{liu2021}. 
Also note that \cite{liu2021} detected the quasar J0227--0605 at $z=6.2$ at 3\,GHz but not at 1.4\,GHz, making it potentially radio-loud, though deeper 1.4\,GHz (or lower frequency) observations are required for a robust classification.}, listed by increasing redshift, are: J0309+2717 at $z=6.10$  \citep{belladitta2020}, J1427+3312 at $z=6.12$ \citep{mcgreer2006,stern2007}, and   J1429+5447 at $z=6.18$ \citep{willott2010a}. 

In this paper we present the discovery and initial characterization of the most distant radio-loud quasar currently known, PSO~J172.3556+18.7734 (hereafter \radioqso) at $z=6.823$, { as measured from the \mgii\, emission line}. In Section \ref{sec:qso} we describe the selection of the quasar and the details of follow-up observations. The properties derived from near-infrared spectroscopy are presented in Section \ref{sec:nir-modeling} and the properties from follow-up radio observations are introduced in Section \ref{sec:radioprops}. We summarize and present our conclusion in Section \ref{sec:remarks}.  Throughout the paper we use a flat cosmology
with $H_0 = 70 \,\mbox{km\,s}^{-1}$\,Mpc$^{-1}$, $\Omega_M = 0.3$, and $\Omega_\Lambda = 0.7$. In this cosmology the age of the universe at the redshift of \radioqso\ is 776\,Myr and 1 pkpc corresponds to 5\farcs3.  Optical and near-infrared magnitudes are reported in the AB system, while for radio observations we report the peak flux density unless otherwise stated. For nondetections we report $3\sigma$ upper limits.

\begin{figure*}[ht]
\plotone{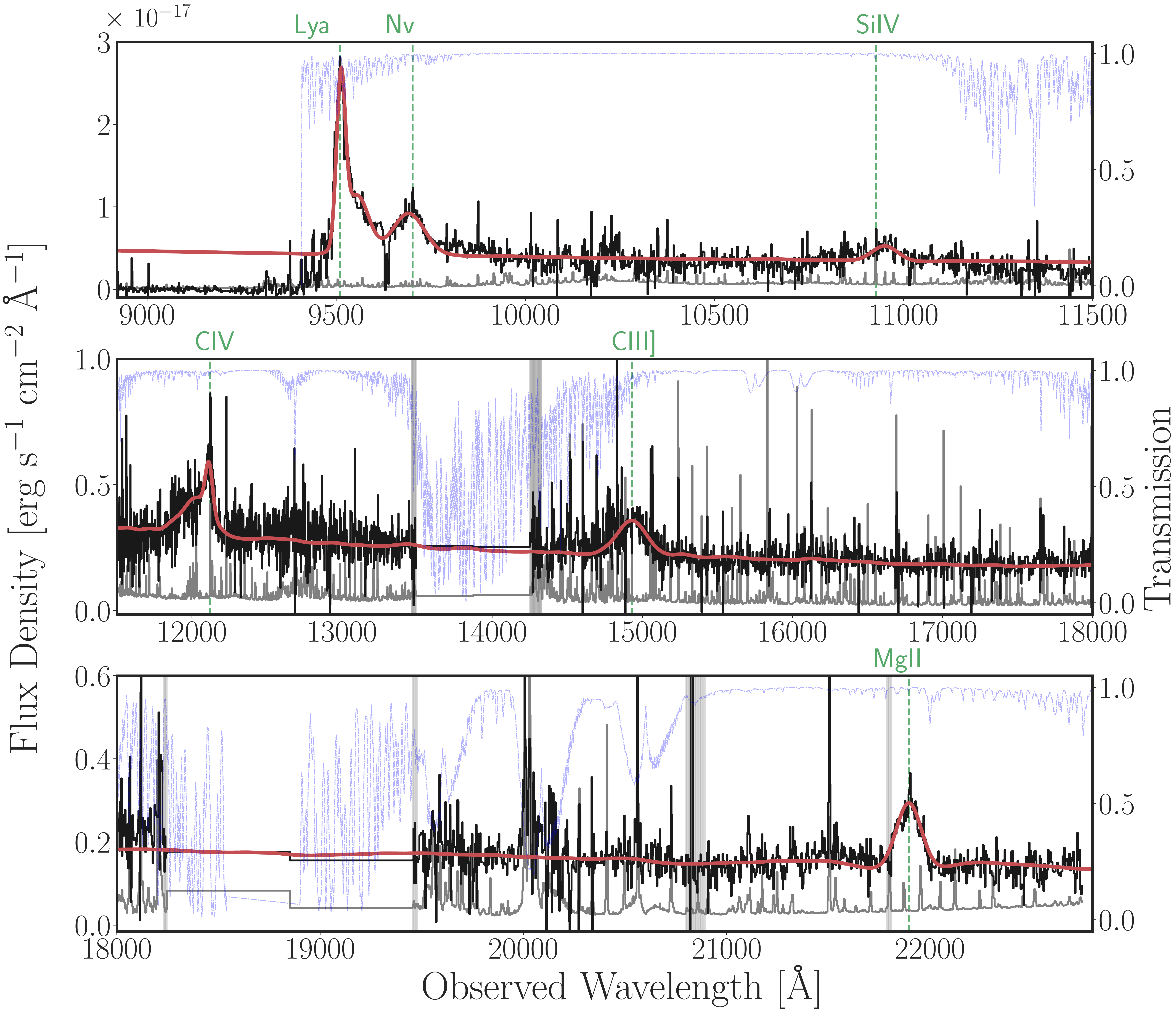}
\caption{
Optical/near-infrared spectrum of P172+18. The final spectrum was obtained by combining all the spectroscopic follow-up data available (Keck/NIRES, VLT/X-Shooter, and LBT/MODS). 
\label{fig:SpecXsh}
}
\end{figure*}

\section{A radio-loud quasar at \texorpdfstring{$\lowercase{z}=6.8$}{z=6.8}} \label{sec:qso}

\subsection{Selection and Discovery}

\radioqso\ has been identified as a $z>6.5$ quasar candidate by at least two independent methods. We first selected \radioqso\  as a $z$-dropout radio-loud candidate in \citealt{banados2015a} (see their Table 1). That selection required red  ($\zps -\yps > 1.4$) sources in the stacked object Pan-STARRS1 catalog \citep{chambers2016} and a counterpart in the radio survey Faint Images of the Radio Sky at Twenty cm  \citep[FIRST,][]{becker1995} to avoid most of the L- and T-dwarfs, which are the main contaminants for $z>6.5$ quasar searches (see \citealt{banados2015a} for details). The Pan-STARRS1 and FIRST measurements for \radioqso\ are listed in Table \ref{tab:photometry}. This object also stands out as a promising high-redshift quasar candidate in a new method to select $z\gtrsim 6.5$ quasars exploiting the overlap of Pan-STARRS1 and the DESI Legacy Imaging Surveys (DECaLS; \citealt{dey2019}), which will be presented in a forthcoming paper along with additional $z\gtrsim 6.5$ quasar discoveries (E.\ Ba\~nados et al.\ in preperation). \radioqso\ was selected using the DECaLS DR7 catalog, but in Table \ref{tab:photometry} we report the photometry from the latest (DR8) data release.

The optical photometry of \radioqso\ in the DECaLS DR7 and DR8 catalogs is consistent. However, the mid-infrared Wide-fied Infrared Survey Explorer (\textit{WISE}) magnitudes are inconsistent\footnote{We also note that \radioqso\ does not appear in the ALLWISE \citep{cutri2014}, unWISE \citep{schlafly2019}, or CatWISE \citep{eisenhardt2020} catalogs.} at the $2\sigma$ level even though DR7 and DR8 use the same input set of \textit{WISE} images spanning from 2010 to 2017 \citep[A.~Meisner, private communication;][]{meisner2019}. 
DECaLS provides matched \textit{WISE} photometry by using the $\gde$, $\rde$, $\zde$ information to infer the \textit{WISE} magnitudes from deep image coadds using all available \textit{WISE} data \citep{lang2014,meisner2017,meisner2019}. The \textit{WISE} DECaLS DR7 magnitudes are $W1=21.25 \pm 0.21$ and $W2=21.30 \pm 0.51$ in contrast to the DR8 magnitudes of $W1=20.71 \pm 0.13$ and $W2=20.73 \pm 0.31$. The main difference between DR7 and DR8 is the change of sky modeling as presented in \cite{schlafly2019}, which can affect the fluxes of sources at the faint limit of the unWISE coadds. Therefore, the reported \textit{WISE} magnitudes need to be taken with caution. 

We confirmed \radioqso\ as a $z\sim 6.8$ quasar on 2019 January 12 with a 450 s spectrum using the Folded-port InfraRed Echellete (FIRE; \citealt{simcoe2008, simcoe2013}) spectrograph in prism mode at the Magellan Baade telescope at Las Campanas Observatory.  The spectrum had poor signal-to-noise  ratio (S/N) but was sufficient to unequivocally identify \radioqso\ as the most distant radio-loud quasar known to date, which triggered a number  of follow-up programs described below.

\subsection{Near-infrared Imaging Follow-up} 

We obtained $JHK$ photometry using the NOTCam instrument at the Nordic Optical Telescope \citep{djupvik2010}.  The total exposure times were 19 minutes each for \Jnot\ and \Hnot\ and 31 minutes for \Knot. Data reduction consisted of standard procedures: bias subtraction, flat-fielding, sky subtraction, alignment, and stacking. Table \ref{tab:obslog} presents a log of the observations.

We calculate the zero-points of the NOT
observations, calibrating against stars in the Two Micron All Sky Survey (2MASS) using the following conversions: 

\begin{eqnarray*}
\Jnot = & \Jtmass - 0.074  \times (\Jtmass - \Htmass) + 0.003\\
\Hnot = & \Htmass + 0.045  \times (\Jtmass - \Htmass) + 0.006\\
\Knot = & \Ktmass + 0.580  \times (\Htmass - \Ktmass) + 0.225\\
\end{eqnarray*}

\noindent  These conversions were calculated via linear fits of the stellar loci as described in Section 2.6 of \cite{banados2014}. The near-infrared photometry is listed in Table~\ref{tab:photometry}.

\begin{figure*}[ht]
\plotone{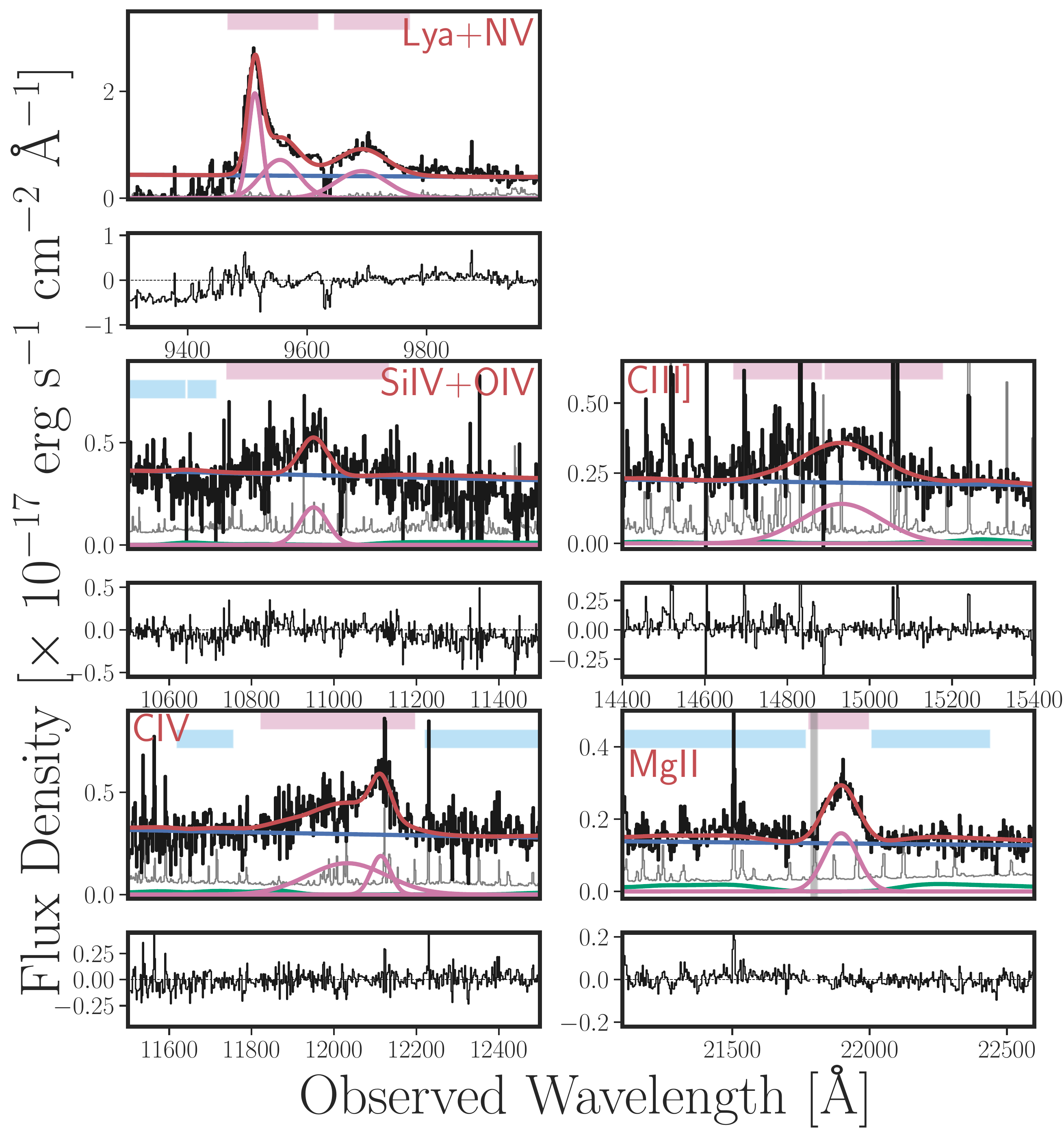}
\caption{
Zoom-in on the main broad emission lines from near-infrared spectroscopy. We show the total spectral fit (red line), and the different components, i.e. power law + Balmer pseudo-continuum (blue line), \feii~template (green line, from \citealt{vestergaard2001}), and emission lines (pink lines). Spectral regions used for the continuum and spectral line fits are shown as horizontal light blue and pink regions, respectively. The noise spectrum is reported in gray in the main panels, while residuals are also shown below each panel. Regions with low S/N or strong absorption features are masked out during the fit, and highlighted with gray vertical regions.
\label{fig:SpecXsh_zoom}
}
\end{figure*}

\begin{figure*}[ht!]
\plotone{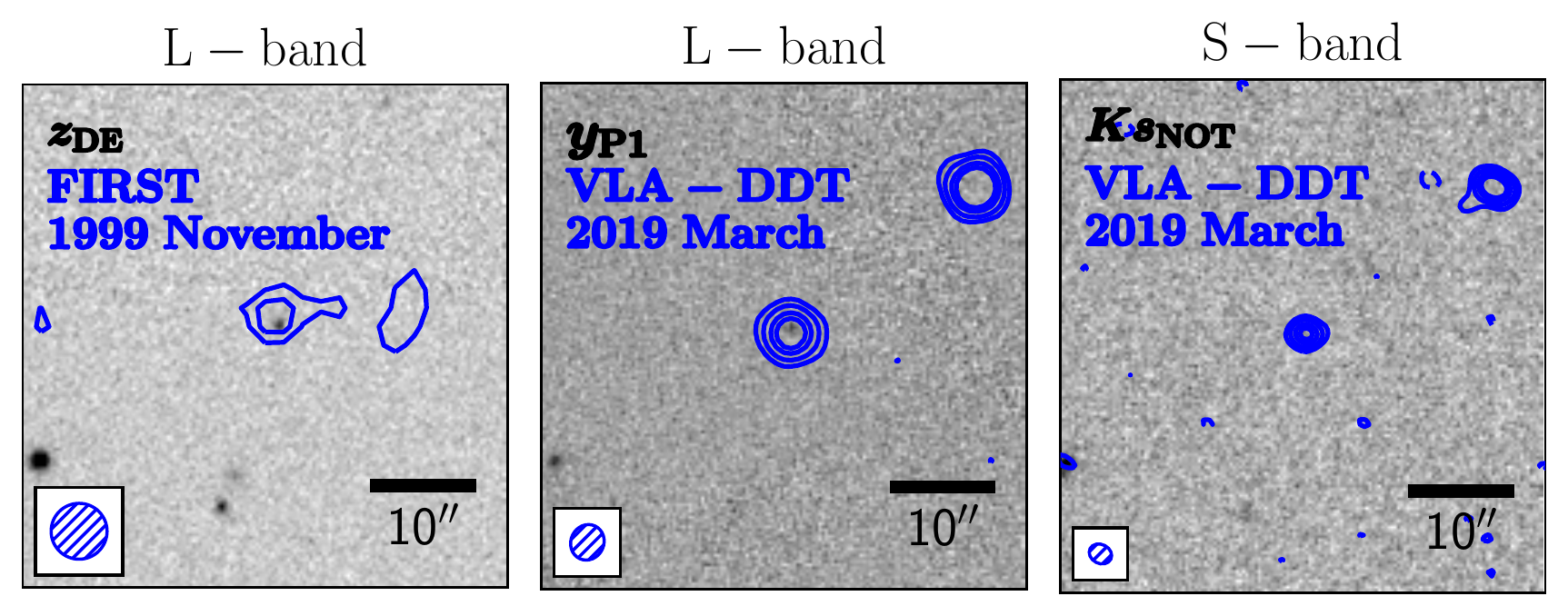}
\caption{
VLA L- and S-band observations (blue contours) centered on the position of \radioqso\ over optical and near-infrared imaging as labeled in the figure; north is up and east is left. 
Contours correspond to 3$\sigma$ and 5$\sigma$ for the FIRST image (left panel) and to 3$\sigma$, 7$\sigma$, 14$\sigma$, and 21$\sigma$ for the VLA-DDT images (corresponding negative contours for all panels are dashed);   $\sigma$ for each of the radio images is listed in Table \ref{tab:obslog}. The follow-up observations reveal a second radio source  23\farcs1 to the northwest of the quasar with no counterpart in available optical or near-infrared imaging (see Table~\ref{tab:photometry}). Although the secondary radio source is slightly brighter than the quasar in the deepest observations, it was not visible in the FIRST survey data. 
     \label{fig:radioimgs}}
\end{figure*}

\subsection{Spectroscopic Follow-up} 
We obtained three follow-up spectra of \radioqso. On 2019 February 18 we observed \radioqso\ for 3.5 hr with Keck/NIRES \citep{wilson2004}. 
Between 2019 March 8 and April 8 we used the Very Lare Telescope (VLT)/X-Shooter spectrograph \citep{vernet2011} to observe the target for a total time of 3.5 hr.  
We also observed \radioqso\ with the Large Binocular Telescope (LBT)/Multi-Object Double Spectrograph (MODS) \citep{pogge2010} on 2019 June 13. 
The MODS observations were carried out in binocular mode 
for 20 minutes on-source. We summarize the spectroscopic follow-up observations in Table~\ref{tab:specobs}.

The Keck/NIRES and VLT/X-Shooter data were reduced with the Python Spectroscopic Data Reduction Pipeline  \citep[PypeIt;][]{prochaska2019,prochaska2020}.
In practice, sky subtraction on the 2D images was obtained through a B-spline fitting procedure and differences between AB dithered exposures. The 1D spectrum was extracted with the optimal spectrum extraction technique \citep{horne1986}. Each 1D single exposure was flux-calibrated using standard stars observed with X-Shooter. Then, the 1D spectra were stacked and a telluric model was fitted, obtained from telluric model grids from the Line-By-Line Radiative Transfer Model (LBLRTM4; \citealt{clough2005}, \citealt{gullikson2014}).
The X-Shooter and NIRES spectra were then absolute-flux-calibrated with respect to the $J_\mathrm{{NOT}}$ magnitude (see Table \ref{tab:photometry}).
The LBT/MODS binocular spectra were reduced with IRAF using standard procedures, including bias subtraction, flat-fielding, and telluric and wavelength calibration. They were each scaled to the \yps\ magnitude.

We performed all measurements presented in the following sections in the individual spectra, which resulted in consistent results. To maximize the information provided by all spectra we re-binned them to a common wavelength grid with a pixel size of  50\,\kms, and averaged them weighting by their inverse variance. 
The final spectrum that we use for our main analysis is shown in Figure $\ref{fig:SpecXsh}$ and a zoom-in on the main emission lines is presented in Figure \ref{fig:SpecXsh_zoom}.

\subsection{Radio Follow-up}
Follow-up radio-frequency observations were carried out with the Karl G.\ Jansky Very Large Array (VLA) of the NRAO\footnote{The National Radio Astronomy Observatory is a facility of the National Science Foundation operated under cooperative agreement by Associated
Universities, Inc.} on 2019 March 5 and 2019 March 11, in S and L bands respectively. Each observing session was 1\,hr in total ($\sim$21 min on-source). The VLA was in B-configuration with a maximum baseline length of 11.1\,km. The observations spanned the frequency ranges 1--2\,GHz (L band; center frequency 1.5\,GHz) and 2--4\,GHz (S band; center frequency 3\,GHz). The WIDAR correlator was configured to deliver 16 adjacent subbands per receiver band, each 64\,MHz at L band and 128\,MHz at S band. Each subband had 64 spectral channels, resulting in 1\,MHz channels in the L-band data and 2\,MHz channels in the S-band data.

The source 3C 286 (J1331+3030) was used to set the absolute flux density scale and to calibrate the bandpass response, and the compact source J1120+1420 was observed as the
complex gain calibrator. Data editing, radio-frequency interference (RFI) excision, calibration, imaging, and analysis were performed using the Common Astronomy Software Applications (CASA) package of the NRAO. The data were calibrated using the CASA pipeline version 5.4.1-23, and the continuum images were made using the wide-field w-projection gridder and Briggs weighting with robust=0.4 as implemented in the CASA task tclean. Due to the excision of data affected by RFI, the resulting L- and S-band images have the reference frequencies of 1.52 and 2.87\,GHz, respectively. 
The resulting beam sizes for the 1.52 and 2.87\,GHz images are $3\farcs55 \times 3\farcs24$  and $2\farcs27 \times 1\farcs85$, respectively.  
A summary of the radio observations is listed in Table \ref{tab:obslog} and the results are discussed in Sections \ref{sec:qso_radio_props} and \ref{sec:comp_radio_props}. 
The follow-up radio images as well as archival data from the FIRST survey are shown in Figure \ref{fig:radioimgs}. 

\section{Analysis of UV--Optical Properties}
\label{sec:nir-modeling}
To derive the properties of the broad emission lines, we use a tool especially designed to model near-infrared spectra of high-redshift quasars, which is described in detail in Section 3 of \cite{schindler2020}. 
Briefly, we consider both the quasar pseudo-continuum emission and the broad emission lines.
In particular, we fit the former with the following components:
\begin{enumerate}
    \item a {\it power law} ($f_{pl}$), normalized at rest-frame wavelength 2500\,$\text{\AA}$:
          \begin{equation}
              f_{pl} = f_{pl,0} \left( \frac{\lambda}{2500\,\ \text{\AA}} \right)^{\alpha_{\lambda}}
          \end{equation}
          where $\alpha_{\lambda}$ and $f_{pl,0}$ are the power-law index and amplitude, respectively.
   
    \item a {\it Balmer pseudo-continuum}. We consider the description from \cite{dietrich2003}, valid for wavelength $\lambda \leq \lambda_{\rm BE} = 3646\,\text{\AA}$, i.e., where the Balmer break occurs:
          \begin{equation}
              f_{BC}(\lambda) = f_{BC,0} B_{\lambda} (\lambda,T_{e}) \left(1- e^{\tau_{\rm BE} (\lambda/\lambda_{\rm BE})^{3}} \right)
          \end{equation}
         with $B_{\lambda} (T_{e})$ the Planck function at electron temperature $T_{e}$, $\tau_{\rm BE}$ the optical depth at the Balmer edge, and $f_{BC,0}$ the normalized flux density at the Balmer break. 
         Following the literature (e.g., \citealt{dietrich2003}, \citealt{kurk2007}, \citealt{derosa2011}, \citealt{mazzucchelli2017b}, \citealt{onoue2020}), we assume $T_{e} = 15,000$\,K and 
         $\tau_{\rm BE} = 1$, and we fix the Balmer emission to 30\% of the power-law contribution at rest-frame $3646\,\text{\AA}$.
   
    \item an {\it \feii~pseudo-continuum}. We model the \feii~contribution with the empirical template from \cite{vestergaard2001},  which is used in the derivation of the scaling relation that we later consider for estimating the black hole mass of the quasar (see Section \ref{sec:qso_spec_prop_bh} and Equation \ref{eq:MBH_VO09}).
    We fit the \feii~in the rest-frame wavelength range 1200 -- 3500\,\AA.
    Assuming that \feii\ emission arises from a region close to that responsible for the \mgii\, emission, we fix $z_{\rm Fe II} = z_{\rm Mg II}$ and FWHM$_{\rm Fe II}$ to be equal to  FWHM$_{\rm Mg II}$.
    
\end{enumerate}
To perform the fit, we choose regions of the quasar continuum free of broad emission lines and of strong spikes from residual atmospheric emission: [1336--1370], [1485--1503], [1562--1626], [2152--2266], [2526--2783], [2813--2869]\,{ \AA} (rest frame).

We subtract the entire pseudo-continuum model (power law + \feii\ + Balmer pseudo-continuum) from the observed spectrum, and then we model the broad emission lines with Gaussian functions, interactively choosing the wavelength range for the fit.
In particular, we model the \nv, \siiv, \ciii, and \mgii~lines with a single Gaussian, while the $\lya$ and \civ\, lines are better fit by two Gaussians representing a narrow component and a broad one. 

After obtaining the best fit,  
we implement a second routine to obtain the best parameters and their uncertainties through a bootstrap resampling approach.
The spectrum is resampled 500 times by drawing from a Gaussian distribution with mean and standard deviation equal to the observed spectrum and the uncertainty on each pixel, respectively.
For every resampling, the spectrum is refit with the initial best fit used as a first guess. 
All the model parameters are then saved and used to build a distribution. 
The final best values and uncertainties correspond to the 50\% and 16\% and 84\% percentiles, respectively. 

We show the total best fit of the final spectrum in Figure \ref{fig:SpecXsh} and zoom-in on the emission lines in Figure \ref{fig:SpecXsh_zoom}. We list the measured quantities in Tables \ref{tab:photometry} and \ref{tab:SpecPropMeas} and the derived properties in Table \ref{tab:props}.
\subsection{Emission Line Properties} \label{sec:qso_spec_prop_el}
Specific properties such as equivalent width (EW) and peak velocity shift of key broad emission lines (e.g.,~$\lya$, \nv, \civ, and \mgii) have been shown to trace properties of the innermost regions of quasars and of their accretion mechanisms (e.g, \citealt{leighly2004}, \citealt{richards2002a,richards2011}).

{ We measure the redshifts of the emission lines as
\begin{equation}
     z_{\rm line} = \frac{\lambda_{\rm line, obs}}{\lambda_{\rm line, rf}} - 1
\end{equation}
where $\lambda_{\rm line,obs}$ is the observed line wavelength, i.e., the peak of the fitted Gaussian function, and $\lambda_{\rm line,rf}$ is the rest-frame line wavelength (see Table \ref{tab:SpecPropMeas}). In case of a line fitted with two Gaussian functions (e.g., \civ\ and $\lya$), we considered the peak wavelength corresponding to the maximum flux value of the full model (see \citealt{schindler2020} for further details).
}

P172+18 presents  strong and narrow $\lya$ and \nv\ emission lines (see Figures \ref{fig:SpecXsh} and \ref{fig:SpecXsh_zoom}). 
We derive the total equivalent equivalent width of $\lya$\,+\,\nv, EW($\lya$\,+\,\nv) $\sim56$\,\AA\ 
(see Table \ref{tab:SpecPropMeas}). This is consistent with the mean of the EW($\lya$\,+\,\nv) distributions for $3<z<5$ and $z>5.6$ quasars as found by \cite{diamond2009}
 and \cite{banados2016}, respectively. 
Notably for a $z\sim7$ quasar, the narrow component of the $\lya$ emission of \radioqso\ can be fitted well by a single Gaussian and there is no evidence for an IGM $\lya$ damping wing (see \citealt{Wang2020ApJ...896...23W}), implying that the surrounding IGM is $>90\%$ ionized (see also Section \ref{sec:qso_near_zon}).

Now we focus on the relation between the \civ\, EW and the blueshift with respect to the \mgii\ line. As a reminder, 
we model the \civ\ line with two Gaussians 
(see Table \ref{tab:SpecPropMeas} and Figure \ref{fig:SpecXsh_zoom}). 
In the following, we consider all the components of the model, i.e., the total line emission\footnote{The properties of the single Gaussian components of the fit of the line are presented in Table \ref{tab:SpecPropMeas}}. We measure \civ\, EW = 21.3$^{+2.4}_{-2.0}$ \AA\,and $\Delta$v$\rm _{MgII - CIV} = 195 \pm 225$ km\,s$^{-1}$. 
In Figure \ref{fig:CIVblueEW}, we place the measurements of P172+18 in the context of quasar populations at $z\sim 2$ and $z\gtrsim 6$. 
For the $z\sim 2$ subsample we select quasars from the Sloan Digital Sky Survey (SDSS) Data Release 7 quasar catalog (DR7; \citealt{shen2011}) using the criteria of  \cite{richards2011}: 
\begin{enumerate}
    \item 1.54$<z<$2.2, to ensure that both \civ\, and \mgii\, emission lines are encompassed by the SDSS spectral wavelength range.
    \item FWHM$\rm _{C IV}$ and FWHM$\rm _{Mg II} >$ 1000 km s$^{-1}$, to select only quasars with broad emission lines. 
    \item FWHM$\rm _{C IV} > 2 \sigma_{FWHM,C IV}$ and EW$\rm _{C IV} > 2\sigma_{EW, C IV}$ and EW$\rm _{C IV} > 5$ \AA, for a reliable fit of the \civ\, line.
    \item FWHM$\rm _{Mg II} > 2 \sigma_{FWHM,Mg II}$ and EW$\rm _{Mg II} > 2\sigma_{EW, Mg II}$, for a reliable fit of the \mgii\, line.
    \item we exclude broad absorption line quasars (\texttt{BAL\_FLAG = 0}).
\end{enumerate}
This yields $22,703$ objects, out of which 1284 are classified as radio-loud with $R_{2500} > 10$. 

As shown in \cite{richards2011}, radio-loud quasars occupy a specific region of the  \civ\, EW--blueshift parameter space:  small blueshifts ($\lesssim 1000\,\kms$) but a wide range of EW values. 
However, note that for each radio-loud quasar several radio-quiet ones with similar rest-frame UV properties can be found, but not necessarily the other way around (Figure~\ref{fig:CIVblueEW}). 
Recently, the \civ\, emission line of $z\gtrsim 6$ quasars has been studied by various researchers (e.g., \citealt{mazzucchelli2017b}, \citealt{meyer2019}). Large blue shifts for these objects are ubiquitous, with median values of $\Delta$v$\rm _{Mg\,II - C\,IV}\sim 1800\,\kms$ \citep{schindler2020} and with extreme values extending to $\Delta$v$\rm _{Mg\,II - C\,IV} \gtrsim\,$5000\,\kms \citep[e.g.,][]{onoue2020}. 
In Figure~\ref{fig:CIVblueEW} we  show the $\Delta$v$\rm _{Mg\,II - C\,IV}$ measurements for $z>6$ quasars from \cite{mazzucchelli2017b}, \cite{shen2019}, \cite{onoue2020}, and \cite{schindler2020}. 

To exclude objects with extremely faint emission lines and/or with spectra with low S/N close to the \civ\, line, we consider only high-$z$ quasars for which EW$_{\rm CIV} > 2 \sigma_{\rm CIV} $ and EW$_{\rm CIV} > 5$\,\AA. 
Out of the three radio-loud quasars at $z>6$ that have near-infrared spectra covering \mgii\ and \civ, only J1429+5447 does not satisfy our criteria owing to its extremely weak emission lines (EW$_{\rm CIV} < 5$\,\AA; \citealt{shen2019}). 
The two radio-loud quasars at $z>6$ in Figure \ref{fig:CIVblueEW}, J1427+3312 and P172+18, show \civ\, emission line properties consistent with what is observed in the radio-loud sample at $z\sim2$. 
A larger sample of radio-loud quasars at high redshift with near-infrared spectra is needed to further investigate whether this trend changes with redshift, and whether the different EW and blueshift properties of radio-loud quasars can inform us about physical properties of their broad-line regions and/or their accretion mode.

\subsection{Black Hole Properties} 
\label{sec:qso_spec_prop_bh}
We compute the quasar bolometric luminosity ($L_{\mathrm{bol}}$) using the bolometric correction presented by \cite{richards2006b}:
   \begin{equation}\label{eq:bolcor}
      L_{\mathrm{bol}} = 5.15\, \lambda \, L_{\lambda} (3000\, \mathrm{\AA}) \, \mathrm{erg\,s^{-1}}
   \end{equation}
where  $L_{\lambda} (3000\, \text{\AA})$ is the monochromatic luminosity at 3000\,\AA\  derived from the power-law model.
We estimate the black hole mass using the \mgii\ line as a proxy through the scaling relation presented by \cite{vestergaard2009}:
   \begin{equation} \label{eq:MBH_VO09}
      M\mathrm{_{BH}} = 10^{6.86} \, \left[ \frac{\mathrm{FWHM (MgII)}}{10^3 \mathrm{km\,s^{-1}}} \right]^{2} \, \left[ \frac{\lambda L_{\lambda} (3000 \,\text{\AA})}{10^{44} \mathrm{erg\,s^{-1}}} \right]^{0.5} \, M_{\odot}.
   \end{equation}
This scaling relation has an intrinsic scatter of 0.55\,dex, which is the dominant uncertainty of the black hole mass estimate. 
Once we have a black hole mass estimate, we can directly derive the Eddington luminosity as 
   \begin{equation}
      L_{\mathrm{Edd}} = 1.3 \times 10^{38} \, \left( \frac{M_{\rm BH}}{M_{\odot}} \right) \, \mathrm{erg\,s^{-1}}.
   \end{equation}

We obtain a black hole mass of $M_{\mathrm{BH}}=2.9^{+0.7}_{-0.6} \times 10^{8} M_{\odot}$ and an Eddington ratio of $L_{\mathrm{bol}}$/$L_{\mathrm{Edd}}=2.2^{+0.6}_{-0.4}$  for \radioqso\ (see also Table~\ref{tab:props}). We note that the Eddington ratio depends on the bolometric luminosity correction used. For example, using the  correction recommended by  \cite{runnoe2012},

   \begin{equation}\label{eq:runnoe2012}
      \log L_{\mathrm{bol}} = 1.852 + 0.975 \times \log (\lambda \, L_{\lambda} (3000\, \text{\AA})),
   \end{equation}

\noindent yields $L_{\rm bol}=6.5\times 10^{46}$\,erg\,s$^{-1}$ and an
Eddington ratio of $L_{\mathrm{bol}}$/$L_{\mathrm{Edd}}=1.8$. For the reminder of the analysis we consider the bolometric correction from equation~\ref{eq:bolcor} to facilitate direct comparison with relevant literature \citep[e.g.,][]{shen2019,schindler2020}. 

\begin{figure}[ht]
\plotone{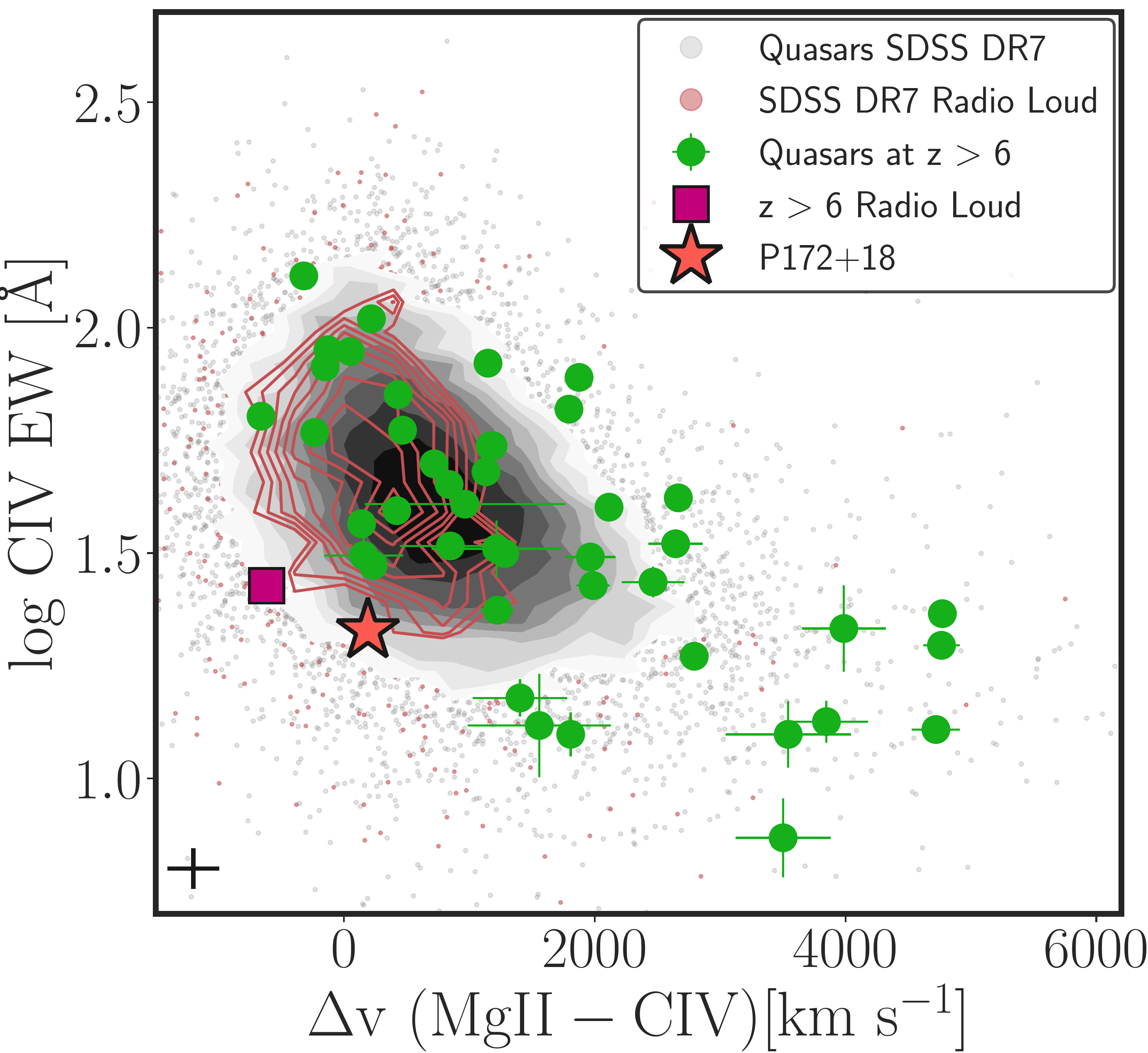}
\caption{
\civ\, equivalent width vs.\ blueshift with respect to the \mgii\, emission line. We show the distribution of   SDSS DR7 $1.4<z<2.2$ quasars with gray points and shaded gray contours. The radio-loud subsample from SDSS is highlighted with red points and contours (see Section \ref{sec:qso_spec_prop_el} for definition and selection). Quasars at $z>6$ are reported with green points, and obtained from a collection of works from the literature (\citealt{mazzucchelli2017b}, \citealt{shen2019}, \citealt{onoue2020}, \citealt{schindler2020}). Before this work, there was only one $z>6$ radio-loud quasar with robust \civ\ and \mgii\ measurements (magenta square). P172+18 at $z=6.823$ is represented as an orange star and its uncertainties are shown in the bottom left corner.
\label{fig:CIVblueEW}
}
\end{figure}

\begin{figure}[ht]
\plotone{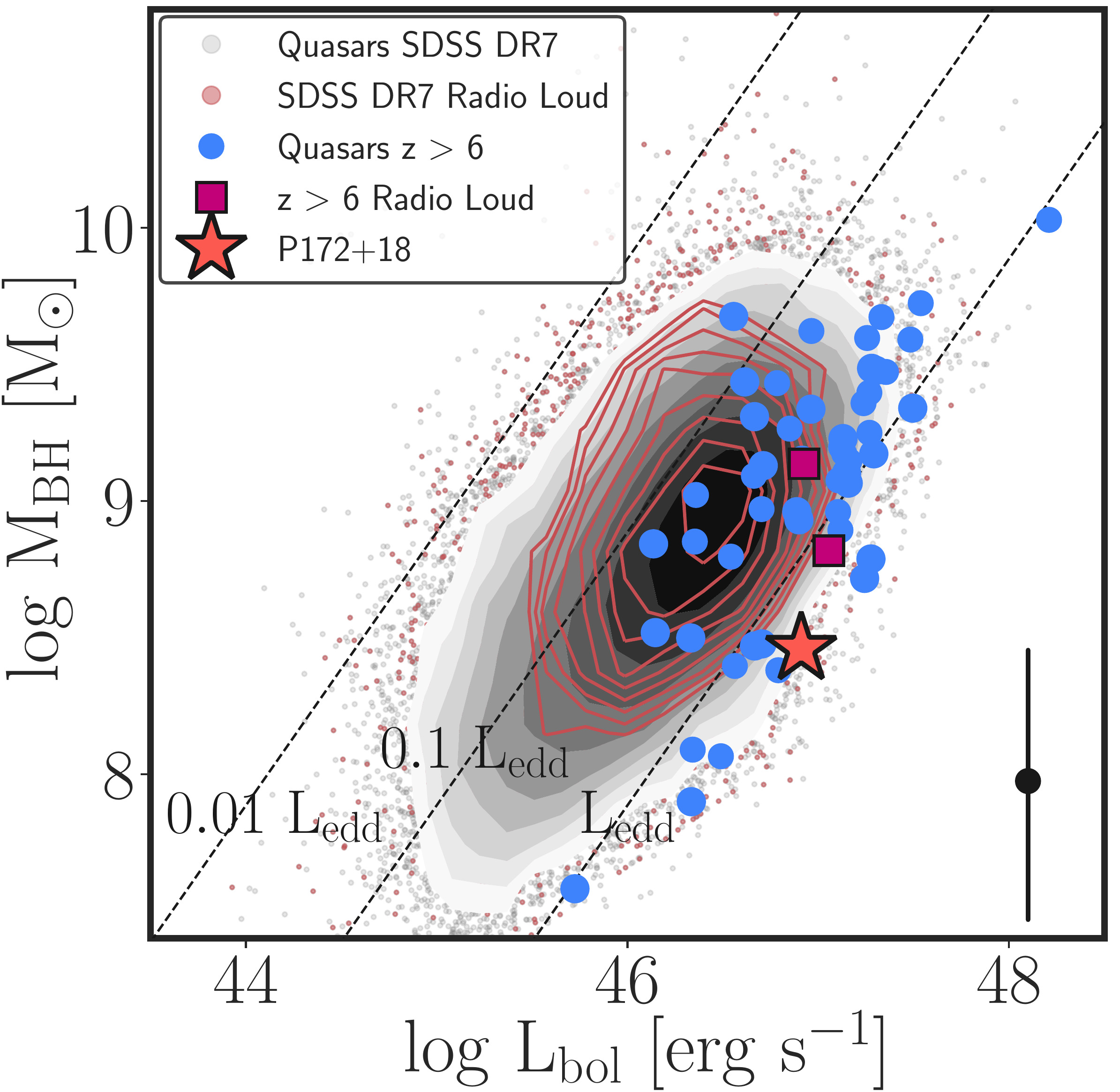}
\caption{
Black hole mass vs.\ bolometric luminosity. The gray points and contours show the distribution of SDSS DR7 quasars at $0.35<z<2.25$. Red points and contours highlight the SDSS DR7 radio-loud quasar subsample. We show $z>6$ radio-quiet and radio-loud quasars from a collection of studies in the literature (see Section \ref{sec:qso_spec_prop_bh} for references) with blue circles and magenta squares, respectively. P172+18 (orange star) is consistent with accreting matter at super-Eddington rate. The dominant systematic uncertainty on black hole mass estimates from scaling relations ($\sim$0.55 dex) is shown in the bottom right corner. All black hole masses shown here are estimated using the same scaling relation \citep{vestergaard2009}, and the same bolometric correction was applied for all bolometric luminosities \citep{richards2006b}.
\label{fig:MBHLbol}
}
\end{figure}

In Figure \ref{fig:MBHLbol} we plot black hole mass vs.\ bolometric luminosity for P172+18 as well as other $z>6$ and lower-redshift quasars from the literature. 
As for Figure \ref{fig:CIVblueEW}, the low-redshift quasar sample is taken from the  SDSS DR7 quasar catalog. Here, we select  objects with redshift $0.35<z<2.25$, i.e., for which the \mgii~emission line falls within the observed wavelength range, and with valid values of FWHM\,(\mgii) and $L_{\lambda}(3000)$, necessary to estimate the black hole masses and bolometric luminosities. This results in $85,504$  SDSS quasars, out of which 5769 are classified  as radio-loud (red contours in Figure \ref{fig:MBHLbol}). 
We compiled the $z>6$ quasar sample from the following studies: \cite{willott2010b}, \cite{derosa2011}, \cite{wu2015}, \cite{mazzucchelli2017b}, \cite{shen2019}, \cite{pons2019}, \cite{reed2019}, \cite{matsuoka2019}, \cite{onoue2019,onoue2020}, and \cite{yang2020}. 
We recalculate the black hole masses and bolometric luminosities of all quasars, at both low and high redshift using equations \ref{eq:bolcor} and \ref{eq:MBH_VO09}.
The two high-redshift radio quasars for which these measurements are available from the literature (J1427+3312 and J1429+5447, both with near-infrared spectra presented by \citealt{shen2019}), show black hole masses and bolometric luminosities consistent with radio-loud quasars at lower redshift, and with the general quasar population at $z>6$.
The black hole of P172+18 is accreting matter at a rate consistent with super-Eddington accretion, and it is found among the fastest accreting quasars at both $z\sim1$ and $z\gtrsim$6. 

\subsection{Near-zone Size} 
\label{sec:qso_near_zon}
Near-zones are regions around quasars where the surrounding intergalactic gas has been ionized by the quasar's UV radiation, and they are observed as regions of enhanced transmitted flux close to the quasar in their rest-frame UV spectra. 
The near-zone sizes fo quasars provide constraints on quasar emission properties and on the  state of their surrounding IGM (e.g.,~\citealt{fan2006a}, \citealt{eilers2017a}, \citeyear{eilers2018a}). 
The radii of near-zones ($R_{\rm NZ}$) depend on the rate of ionizing flux from the central source, on the quasar's lifetime, and on the ionized fraction of the IGM (e.g., \citealt{fan2006a}, \citealt{davies2019}). 
In practice, $R_{\rm NZ}$ is measured from the rest-frame UV spectrum (smoothed to a resolution of 20\,\AA) and taken to be the distance from the quasar at which the transmitted continuum-normalized flux drops below 10\%. 
Here, we obtain the transmitted flux by dividing the observed spectrum of P172+18 by a model of the intrinsic continuum emission obtained with a principal components analysis method (see \citealt{davies2018a}; \citealt{eilers2020a}, for details of the method). In order to take into account the dependence on the quasar's luminosity, we also calculate the corrected near-zone radius ($R_{\rm NZ,corr}$), following the scaling relation presented by \cite{eilers2017a}:
\begin{equation}
  R_{\rm NZ,corr} = R_{\rm NZ} \times 10^{0.4(27+M_{1450})/2.35}    
\end{equation}
where $M_{1450}$ is the absolute magnitude at rest-frame 1450\,\AA. We report both $R_{\rm NZ}$ and $R_{\rm NZ,corr}$ in Table~\ref{tab:props}. The size of the near-zone of P172+18 and the corrected near-zone  are $R_{\rm NZ}=3.96\pm0.48$~pMpc and $R_{\rm NZ,corr}=6.31\pm0.76$~pMpc, respectively. This large near-zone is within the top quintile of the distribution of quasar near-zones at $z\gtrsim 6$ \citep{eilers2017a}. This suggests that the time during which this quasar is UV-luminous (here referred to its lifetime) exceeds the average lifetime of the high-redshift quasar population of $t_{\rm Q}\sim10^{6}$~yr \citep{eilers2020a}. 

The evolution of $R_{\rm NZ,corr}$ with redshift, at $z>5.5$, has been investigated in the literature to constrain both the reionization history and quasar lifetimes (e.g., \citealt{carilli2010},  \citealt{davies2016a}, \citealt{eilers2020a}). While \cite{carilli2010} and \cite{venemans2015} recover a steep decline of $R_{\rm NZ, corr}$ with redshift (a decrease in size by a factor of  $\sim 6$  between $z=6$ and $z=7$), \cite{eilers2017a} study a  larger sample of $\sim$30 quasars at $5.8<z<6.6$ and recover a best-fit relation in the form of $R_{\rm NZ,corr} \propto (1+z)^{-\gamma}$, with $\gamma\sim1.44$, suggesting a more moderate evolution with redshift than previous studies (a reduction in size by only $\sim 20\%$ between $z=6$ and $z=7$). 
Finally, \cite{mazzucchelli2017b} recover a flatter relation ($\gamma\sim1.0$), utilizing measurements of $R_{\rm NZ,corr}$ up to $z\sim7$ (see also \citealt{ishimoto2020}). 
Using hydrodynamical simulations, \cite{chen2020} obtained a shallow redshift evolution of near-zone sizes over the redshift range probed by the current quasar sample, i.e., $5.5< z<7$ \citep[see also][]{davies2020}. The expected average corrected near-zone size at $z=6.8$ is $\langle R_{\rm NZ,corr}\rangle \approx 4.2$~pMpc for the  redshift evolution from \citet{eilers2017a}, and $\langle R_{\rm NZ,corr}\rangle \approx 2.2$~pMpc when assuming a steeper evolution as found by \citet{venemans2015}. 

Therefore, the new near-zone measurement for P172+18 is considerably larger than the expected average size at this redshift. 
However, if the quasar was more luminous in the recent past and its activity is currently in a receding phase (see \S~\ref{sec:qso_radio_props} for tentative evidence of a decrease in the quasar's radio luminosity), the large near-zone size could be explained by a higher luminosity than what is  measured at the present time.

\begin{deluxetable}{lrr}
\tablecaption{Photometry of the radio-loud quasar \radioqso\ and its radio companion.}
\tablehead{\colhead{} &\colhead{Quasar} &\colhead{Radio Companion
\label{tab:photometry}} 
} 
\decimals 
\startdata 
R.A.\ (J2000) & $11^{\rm h} 29^{\rm m} 25\fs37$  & $11^{\rm h} 29^{\rm m} 24\fs08$  \\
Decl.\ (J2000) & $+18^{\circ} 46^{\prime} 24 \farcs 29$ & $+18^{\circ} 46^{\prime} 38 \farcs 58$\\
\tableline
\multicolumn{3}{c}{Public optical and infrared surveys}\\
Pan-STARRS1 $\ips$ & $>23.6$          & $>23.6$ \\
Pan-STARRS1 $\zps$ & $>23.2$          & $>23.2$ \\
Pan-STARRS1 $\yps$ & $20.76 \pm 0.09$ & $>22.3$ \\
DECaLS DR8 $\gde$  & $>25.4$          & $>25.4$ \\
DECaLS DR8 $\rde$  & $>24.8$          & $>24.8$ \\
DECaLS DR8 $\zde$  & $21.64 \pm 0.05$ & $>23.8$ \\
DECaLS DR8 $W1$          & $20.71 \pm 0.13$ & $>21.8$ \\
DECaLS DR8 $W2$          & $20.73 \pm 0.31$ & $>20.9$ \\
\tableline
\multicolumn{3}{c}{Follow-up near-infrared imaging}\\
$\Jnot$ & $20.90 \pm 0.11$ & $>22.2$\\
$\Hnot$ & $21.36 \pm 0.24$ & $>21.8$\\
$\Knot$ & $21.07 \pm 0.18$ & $>21.7$\\
\tableline
\multicolumn{3}{c}{Public radio surveys}\\
TGSS 147.5\,MHz    & $<8.5\,\mathrm{mJy}$ & $<8.5\,\mathrm{mJy}$\\
 FIRST 1.4\,GHz    & $1020 \pm 144 \,\mu\mathrm{Jy}$\tablenotemark{a} & $<406\,\mu\mathrm{Jy}$\\
 \tableline
 \multicolumn{3}{c}{Radio follow-up}\\
 VLA-L $1.52\,$GHz  & $510 \pm 15 \, \mu\mathrm{Jy}$ & $732 \pm 15 \, \mu\mathrm{Jy}$ \\
 VLA-S $2.87\,$GHz  & $222 \pm 9 \, \mu\mathrm{Jy}$ & $432 \pm 20 \, \mu\mathrm{Jy}$\tablenotemark{b}\\
 $\alpha_{S}^{L}$ & $-1.31 \pm 0.08$ & $-0.83 \pm 0.08$ \\
 \tableline
 \tableline
  \multicolumn{3}{c}{Quasar rest-frame luminosities\tablenotemark{c}}\\
$m_{1450}$  & $21.08 \pm 0.10$ &  \\
$M_{1450}$  &  $-25.81\pm 0.10$ & \\
$L_{2500}$  & $(1.4 \pm 0.1) \times 10^{46}\,$erg\,s$^{-1}$ &   \\
$L_{3000}$  & $(1.3 \pm 0.1) \times 10^{46}$ erg\,s$^{-1}$ &  \\
$L_{4400}$ & $(1.1 \pm 0.1) \times 10^{46}\,$erg\,s$^{-1}$  & \\
$L_{5\,\mathrm{GHz}}$ & $(5.4 \pm 0.2) \times 10^{42}\,$erg\,s$^{-1}$ \\ 
 \enddata 
 \tablenotetext{a}{\footnotesize This is the reported peak flux density in the FIRST catalog (version 2014dec17). 
 We note that in the FIRST image we measure  $852\pm 135\,\mu\mathrm{Jy}$.} 
 \tablenotetext{b}{\footnotesize The source is marginally resolved in the VLA-S image and we report the integrated flux.}
  \tablenotetext{c}{\footnotesize The quasar UV and optical luminosities are derived from the best-fit power law of the near-infrared spectrum (see Table~\ref{tab:SpecPropMeas}) and the uncertainties are dominated by the \Jnot\ photometry used for absolute flux calibration of the spectrum. The 5\,GHz radio luminosity is extrapolated using the measured  radio index.}
\end{deluxetable} 

\section{Analysis of Radio Properties}
\label{sec:radioprops}

In addition to detecting the quasar, the follow-up VLA radio observations revealed a second radio source 23\farcs1 from \radioqso\ at a position angle of $128.25^\circ$ (see Figure~\ref{fig:radioimgs}). 
We will explore the radio properties of the quasar and the serendipitous companion radio source below.

\subsection{Quasar Radio Properties} \label{sec:qso_radio_props}

The quasar is a point source in both the follow-up L- and S-band observations with a deconvolved size smaller than $1\farcs9 \times 0\farcs87$; see Figure \ref{fig:radioimgs}.  \radioqso\ is well detected in both bands with S/N\,$>20$ and the measured flux densities are listed in Table \ref{tab:photometry}. The measured L-band flux density is a factor of two fainter than what is reported in the FIRST catalog. In fact, the  measured $f_{1.52\,\mathrm{GHz}}=510\pm 15\,\mu$Jy would have been below the detection threshold of the FIRST survey \citep{becker1995}. The discrepancy is significant at more than $3\sigma$ and could be the result of real quasar variability over the 20 yr ($\sim 2.5$ yr rest frame)  between the two measurements; such changes have been reported in similar timescales \cite[e.g.,][]{nyland2020}. However, given that the source is at the faint limit of the FIRST survey, we cannot rule out that the variation is simply due to noise fluctuations in the FIRST data.  Unfortunately, we are not able to test the variability hypothesis given that no other measurements of the quasar are available at a similar epoch to the FIRST observation. 
For the remainder of the analysis we will consider the follow-up VLA measurements as the true fluxes. Assuming that the radio observations follow a power-law spectral energy distribution ($f_\nu \propto \nu^{\alpha}$), the L- and S-band flux densities correspond to a steep power-law radio slope of $\alpha_S^L = -1.31 \pm 0.08$. This is steeper than $\alpha=-0.75$, which is  usually assumed in high-redshift quasar studies when only one radio band is available (e.g., \citealt{wang2007, momjian2014, banados2015a}).

\subsubsection{Radio-loudness} \label{sec:radio-loudness}
To estimate the radio-loudness of \radioqso\ we obtain the rest-frame 5\,GHz emission by extrapolating the radio emission using the measured spectral index $\alpha_S^L = -1.31$ and the 2500\,\AA\ and 4400\,\AA\ emission using the power-law fit to the near-infrared spectrum ($\alpha_{\nu,\mathrm{UV}} = -0.48$) obtained in Section \ref{sec:nir-modeling}. This results in radio-loudness parameters of $R_{2500}=91\pm 9$ and $R_{4400}=70\pm 7$, classifying \radioqso\ as a radio-loud quasar. The quasar radio properties are summarized in Table \ref{tab:props}.

\begin{figure}[ht]
\plotone{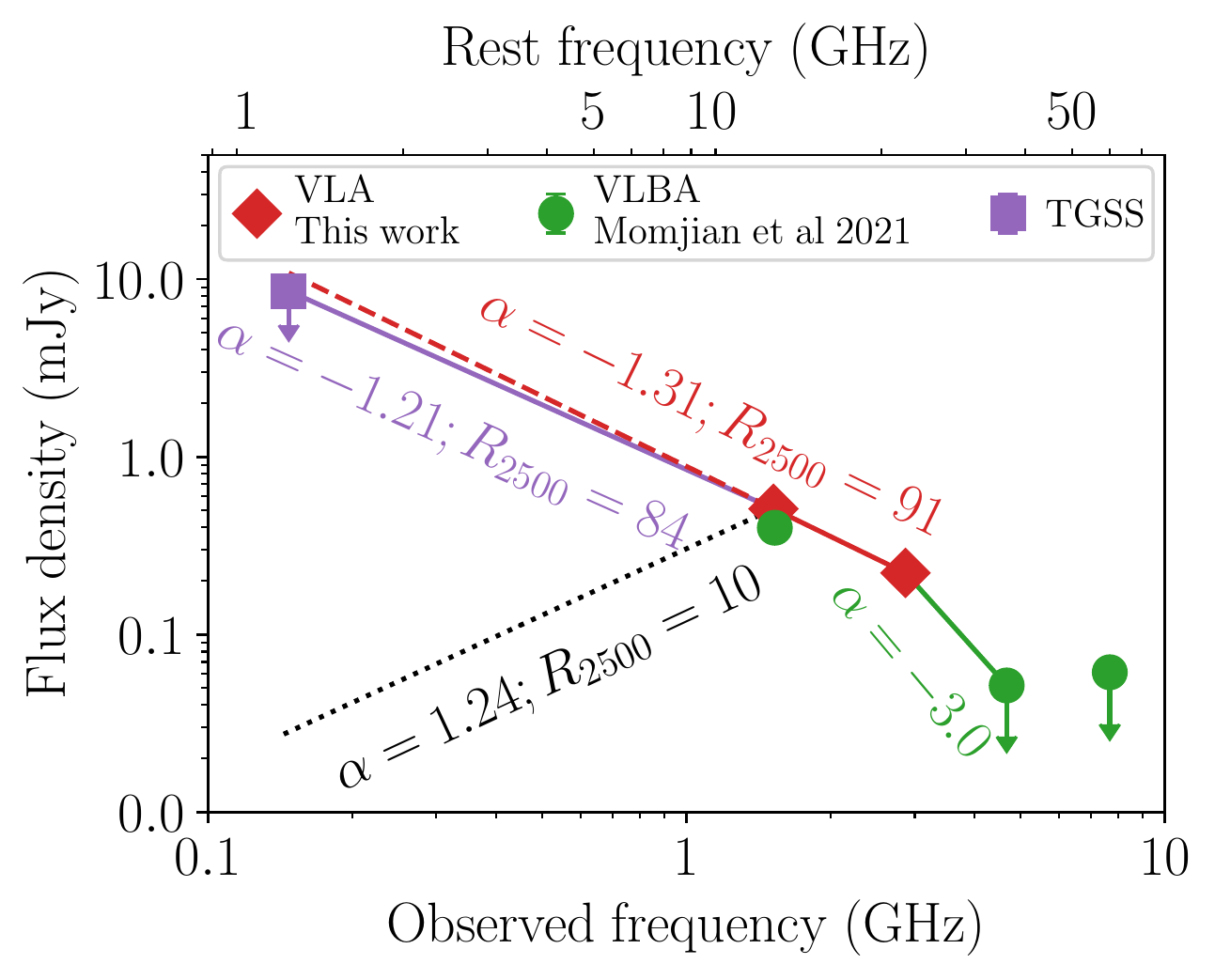}
\caption{
Radio spectral energy distribution of \radioqso, including data from our VLA follow-up observations (red diamond), the VLBI measurements (green circles) from \cite{momjian2021}, and a $3\sigma$ upper limit from the TGSS (purple square). The power-law index, $\alpha$, is shown between the measurements as well as the radio-loudness by extrapolating the radio emission to rest-frame 5GHz. The dotted line with $\alpha=1.24$ represents the turnover required for \radioqso\ to be classified as radio-quiet (i.e., $R_{2500}<10$). 
\label{fig:SED}
}
\end{figure}

We note that the data from very long baseline interferometry (VLBI) presented by \cite{momjian2021} imply a steeper spectral index at frequencies higher than 3\,GHz (see Figure \ref{fig:SED}). The quasar is not detected in the TIFR GMRT Sky Survey \citep[TGSS;][]{intema2017} at 147.5\,MHz. We downloaded the TGSS image and determined a $3\sigma$ upper limit of 8.5\,mJy (see Table \ref{tab:photometry} and Figure \ref{fig:SED}). This implies that the slope of the radio spectrum should flatten or have a turnover between 147.5\,MHz and 1.52\,GHz. If the turnover occurs at a frequency larger than rest-frame 5\,GHz, the rest-frame 5\,GHz luminosity (and therefore radio-loudness) would be smaller than our fiducial value assuming $\alpha=-1.31$. In the extreme case that the turnover happened exactly at the frequency of our L-band observations, the source would still be classified as radio-loud (i.e., $R_{2500}>10$) as long as $\alpha<1.24$ (see Figure \ref{fig:SED}). Deep radio observations at frequencies $<1$\,GHz  are needed to precisely determine the rest-frame 5\,GHz luminosity and the shape of the radio spectrum.

\subsection{Companion Radio Source}  \label{sec:comp_radio_props}

The radio companion is detected with S/N\,$>$\,$20$ in both L- and S-band observations (see Figure \ref{fig:radioimgs}). This object is  a point source in the L-band image with a deconvolved size smaller than $1\farcs6 \times 0\farcs69$. A Gaussian fit to the S-band image results in a resolved source with a deconvolved  size of $1\farcs3 \times 0\farcs8$ and position angle of $74^\circ \pm 22^\circ$.  This secondary source is not detected in any of our available optical, near-infrared, and mid-infrared images. Its radio properties and optical/near-infrared limits are listed in Table \ref{tab:photometry}.

The number of radio sources with a 1.4\,GHz flux density $>700\,\mu$Jy is 59\,deg$^{-2}$ and 117\,deg$^{-2}$ according to the number counts of deep radio surveys from \cite{fomalont2006} and \cite{bondi2008}, respectively. This means that in an area encompassing the quasar and the second radio source ($\pi \times 23\farcs1^2$) only 0.007 and 0.015 sources like the companion are expected using the number counts from \citet{fomalont2006} and \cite{bondi2008}, respectively. The $<2\%$ likelihood of chance superposition raises the possibility that this radio source and the quasar could be associated.

This companion radio source is (slightly) brighter than the quasar in both the  L- and S-band follow-up observations. However, it was not detected in the FIRST survey carried out in 1999 (see Table \ref{tab:photometry} and Figure \ref{fig:radioimgs}). 
This second source could not be a hot spot of the radio jet expanding for the last 20 yr: at the redshift of the quasar, the projected separation of the two sources is about 120 proper kpc, a distance that would take light about $400,000$ yr to travel.   

Another possibility is that this second source is an obscured, radio-AGN companion. There are a few examples of associated dust-obscured, star-forming companion galaxies to quasars at $z>6$ \citep[e.g.,][]{decarli2017, neeleman2019}. A couple of them have tentative X-ray detections, which make them obscured AGN candidates \citep[e.g.,][]{connor2019,vito2019a}. This possibility is tempting, because two associated radio-loud AGNs  would point to an overdense environment in the early universe and provide constraints on AGN clustering. Nevertheless,  with the available shallow optical and near-infrared data we are not able to rule out that the second radio source lies at a different redshift than that of the quasar. More follow-up observations are required to firmly establish the nature and redshift of the source. 

\begin{figure*}[ht]
\plotone{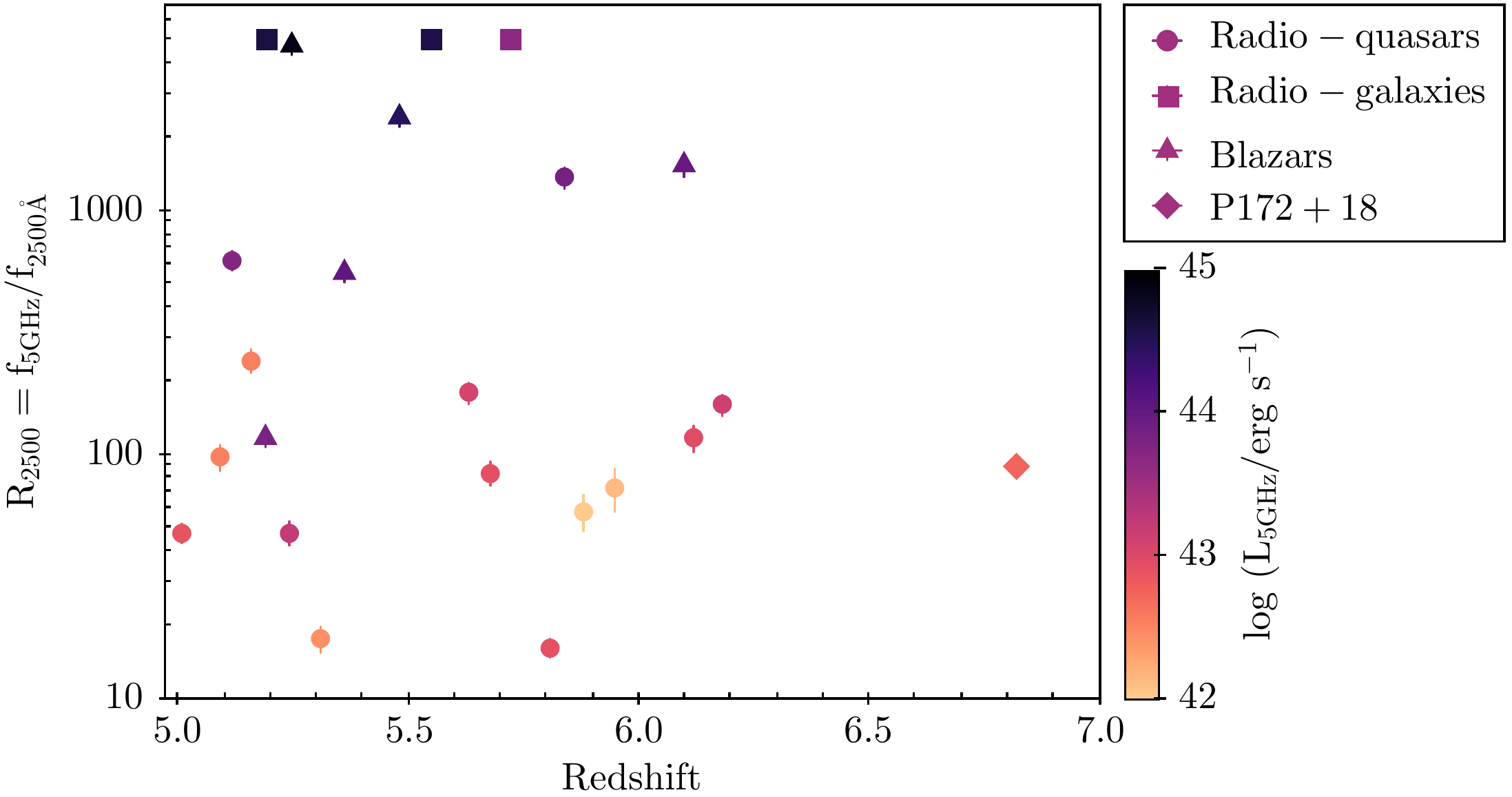}
\caption{
All radio-loud ($R_{2500}> 10$) sources known at $z>5$, color-coded by their rest-frame 5\,GHz radio luminosity. Radio-loud quasars are shown as circles and radio galaxies as squares.  
 The radio-loudness for radio galaxies has been fixed to 5000 for visualization purposes. The properties and references for all these sources are listed in Table~\ref{tab:radio-census}.
 \label{fig:radio_dist}
}
\end{figure*}

\section{Summary and conclusions} \label{sec:remarks}

The main results of this work can be summarized as follows. 

\begin{enumerate}
    \item We present the discovery of the most distant radio-loud source to date, the quasar \radioqso\ with an \mgii-based redshift of  $z=6.823$ (see Figures \ref{fig:SpecXsh} and \ref{fig:radio_dist}  and Table \ref{tab:props}).
    
    \item The \civ\ properties of the two $z>6$ radio-loud quasars known with near-infrared spectroscopy and reliable \civ\ detection (J1427+5447 and \radioqso) are consistent with the radio-loud population at $z\sim 2$ in terms of \civ\ EW and blueshift (see Figure \ref{fig:CIVblueEW}). 
    
    \item The quasar has a black hole mass of $\sim 2.9 \times 10^{8} M_{\odot}$ and an Eddington ratio of $\sim$2.2. It is known that there are large uncertainties on the estimates of black hole mass and Eddington ratio associated with the scaling relations used. Therefore we compare the properties of \radioqso\ to other quasars using the same scaling relation \citep{vestergaard2009} and bolometric correction \citep{richards2006b}. With this in mind,  P172+18 is among the fastest accreting quasars at both low and high redshift (Figure \ref{fig:MBHLbol}). 
    
    \item The quasar shows a strong $\lya$ line that can be modeled with a narrow Gaussian and a broad one  (see Figure \ref{fig:SpecXsh_zoom} and Table \ref{tab:specobs}). The large measured near-zone size, $R_{\rm NZ,corr}\sim 6$\,pMpc, suggests an ionized IGM around the quasar and  implies that \radioqso's lifetime exceeds the average lifetime of the $z\gtrsim 6$ quasar population (see Section \ref{sec:qso_near_zon}). 
    
    \item The quasar's radio emission is unresolved  (with size smaller than $1\farcs90\times0\farcs87$) and shows a steep radio spectrum  ($\alpha =-1.31 \pm 0.08$) between 1.5 and 3.0\,GHz ($\sim$11--23\,GHz in the rest frame). Extrapolating the spectrum to 5\,GHz rest frame, the quasar has a radio-loudness of $R_{2500}=91\pm 9$ (see Figure \ref{fig:SED}). 
    
    \item The follow-up L-band radio data are a factor $\sim 2$ fainter than what is expected from the FIRST observations taken two decades previously.  This fact, together with the  long lifetime implied by  the size of \radioqso's near-zone, could indicate that we are witnessing the quasar phase turning off.

    \item The VLA follow-up observations revealed a second radio source 23\farcs1 from the quasar with comparable radio flux densities (see Figure \ref{fig:radioimgs} and Table \ref{tab:photometry}). This source was not detected in the FIRST survey and has no counterpart in our current optical/near-infrared images. 
\end{enumerate}

 \radioqso, in particular, is an ideal target to investigate the existence of extended X-ray emission arising from the interaction between relativistic particles in radio jets and a hot cosmic microwave background (CMB) \citep[e.g.,][]{Wu2017MNRAS.468..109W}. This effect is expected to be particularly strong at the highest redshifts because the CMB energy density scales as $(1+z)^4$ and as a result its effective magnetic field can be stronger than the one in radio-lobes \citep{ghisellini2015}. Complementary to this science case will be high-resolution VLBI observations to constrain the structure of the radio emission \citep[e.g.,][]{frey2008,momjian2008,momjian2018}. VLBI observations for \radioqso\ already exist and the results will be presented in the companion paper by \cite{momjian2021}.

 The serendipitous detection of the companion radio source (see Figure \ref{fig:radioimgs}) deserves further follow-up. If the radio source lies at the same redshift as the quasar, this could be the most distant AGN pair known, potentially revealing a very dense region in the early universe. Telescopes such as the Atacama Large Millimeter/submillimeter Array or  the James Webb Space Telescope should be able to determine the exact redshift by identifying far-infrared and optical emission lines from this possible obscured AGN. 
 
Out of the 18 quasars known at $z>6.8$, \radioqso\ is the only one currently classified as radio-loud. In Table~\ref{tab:radio-census} we compile the information on all the radio-loud sources at $z>5$ known to date and in Figure~\ref{fig:radio_dist} we present their redshift and radio-loudness distribution. 
The radio-loudness of \radioqso\ is consistent with the median value of the  $z>5$ radio-loud quasar population ($R_{2500,\mathrm{median}}=91$; $R_{2500,\mathrm{mean}}=213$).  
 Thus, the existence of this ``median'' radio-loud quasar at $z=6.823$ makes it likely that there are other radio-loud sources waiting to be discovered (or categorized) between this redshift and the previous redshift record, and possibly even at $z>7$. Identifying these radio sources would be important for future 21cm absorption studies of the IGM with the Square Kilometre Array \citep{carilli2004a,carilli2004c,ciardi2015}.

\begin{deluxetable*}{lcccc}
\tablecaption{Observing Log of Optical and Radio Imaging\label{tab:obslog}}
\tablewidth{0pt}
\tablehead{
\colhead{Date} & \colhead{Telescope/Instrument} & \colhead{Filters/frequency} & \colhead{rms ($1\sigma$)}  & \colhead{Reference}
}
\startdata
2019 May 17 & NOT/NOTCAM & $\Jnot$,$\Hnot$ & 23.4, 23.0 mag & This work\\
2019 May 18 & NOT/NOTCAM & $\Knot$ & 22.9 mag  &This work \\
\hline
1999 Nov 10 & VLA/L-band & 1.44\,GHz & 135\,$\mu$Jy & FIRST   \\
2019 Mar 5 & VLA/S-band & 2.87\,GHz & 9\,$\mu$Jy & This work   \\
2019 Mar 11 & VLA/L-band & 1.58\,GHz & 15\,$\mu$Jy & This work   \\
\enddata
\end{deluxetable*}

\begin{deluxetable*}{lcccc}
\tablecaption{Summary of the optical and near-infrared follow-up spectroscopic observations.\label{tab:specobs}}
\tablewidth{0pt}
\tablehead{
\colhead{Date} & \colhead{Telescope/Instrument} & \colhead{Exposure Time} & \colhead{Wavelength Range} & \colhead{Slit Width} 
}
\startdata
2019 Feb 18 & Keck/NIRES & 3.5 hr & 9400$-$24000 \AA & 0\farcs55\\
2019 Mar 8$-$Apr 8 & VLT/X-Shooter & 3.5 hr & 3000$-$24800 \AA & 1\farcs0 / 0\farcs9 / 0\farcs6 \\
2019 Jun 13 & LBT/MODS & 0.3 hr & 5000$-$10000 \AA & 1\farcs22 
\enddata
\end{deluxetable*}

\begin{deluxetable*}{lcccc}
\tablecaption{Properties of \radioqso\ measured from near-infrared spectroscopy. \label{tab:SpecPropMeas}}
\tablewidth{0pt}
\tablehead{
\colhead{Emission Line} & \colhead{Redshift} & \colhead{FWHM} & \colhead{EW}  & \colhead{$\rm \Delta$v$\rm _{MgII-line}$} \\
 & & \colhead{(km s$^{-1}$)} & \colhead{($\rm \AA$)} & \colhead{(km s$^{-1}$)}
}
\startdata
$\lya$\, (1) & 6.8234$\pm$0.0002 & 891$^{+27}_{-25}$ & -- & $-$15$\pm$77 \\
$\lya$\, (2) & 6.854$\pm$0.001 & 2870$^{+91}_{-77}$ & -- & $-$304$\pm$88 \\
$\lya$ \tablenotemark{a} & 6.8246$\pm$0.0008 & 1103$^{+27}_{-22}$ & 38.1$\pm$1.8 & $-$60$\pm$83 \\
\nv & 6.817$\pm$0.001 & 3076$^{+0.3}_{-0.1}$ & 18.1$\pm$1.1 & 215$\pm$91 \\
$\lya$ + \nv & -- & -- & 56.3$\pm$3.0 & --\\
\siiv + \oiv & 6.822$\pm$0.05 & 3044$^{+1104}_{-652}$ & 8.6$^{+1.7}_{-1.4}$ & 38$\pm$1918   \\
\civ\, (1)  & 6.819$^{+0.004}_{-0.007}$ & 1699$^{+840}_{-499}$ & -- & 153$\pm$224 \\
\civ\, (2)  & 6.753$^{+0.02}_{-0.03}$  & 5001$^{+1018}_{-2143}$ & -- & 2682$\pm$961\\  
\civ \tablenotemark{a} & 6.818$^{+0.004}_{-0.006}$ & 2714$^{+1704}_{-945}$ & 21.3$^{+2.4}_{-2.0}$ & 192$\pm$225 \\
\ciii    & 6.799$^{+0.008}_{-0.005}$ & 5073$^{+560}_{-461}$ & 28.9$^{+3.4}_{-3.3}$ & 920$\pm$260   \\
\mgii \tablenotemark{b}    & 6.823$^{+0.003}_{-0.001}$ & 1780$^{+100}_{-50}$ & 20.8$^{+2.8}_{-2.6}$ & -- \\
\hline
Power-law slope ($\alpha_{\lambda,\mathrm{UV}}$) & --1.52 $\pm$ 0.05 \\ 
Power-law ampl.\ ($f_{\lambda,2500\mathrm{,obs}}$)  & 1.36 $\pm$ 0.03 [$\times$ 10$^{-18}$ erg s$^{-1}$ cm$^{-2}$\,\text{\AA}$^{-1}$]\\ 
\hline
\enddata
\tablenotemark{b}{\footnotesize  The redshift of P172+18 used throughout the paper is taken from the fit to the \mgii\ emission line, as reported here.}
\end{deluxetable*}

\begin{deluxetable*}{lc}
\tablecaption{\radioqso\ properties derived from the optical and near-infrared spectroscopy and radio observations. \label{tab:props}}
\tablewidth{0pt}
\tablehead{
\colhead{Quantity} &  \colhead{ }
}
\startdata
$L_{\mathrm{bol}}$ & (8.1 $\pm 0.3) \times 10^{46}$ erg s$^{-1}$ \\
$M_{\mathrm{BH}}$  & 2.9$^{+0.7}_{-0.6} \times 10^{8}$ $M_{\odot}$ \\ 
$L_{\mathrm{bol}}$/$L_{\mathrm{Edd}}$ & 2.2$^{+0.6}_{-0.4}$ \\ 
\feii/\mgii\ & 2.8$\pm$1.0 \\ 
$R_{\rm NZ}$ & $3.96\pm0.48~\rm pMpc$\\
$R_{\rm NZ,corr}$ & $6.31\pm0.76~\rm pMpc$\\
$R_{2500}$ & $91 \pm 9$ \\
$R_{4400}$ & $70 \pm 7$ \\ 
\hline
\enddata
\end{deluxetable*}

\begin{deluxetable*}{lcccccccc}
\tablecaption{Census and Properties of $z>5$ Radio-loud Sources, Sorted by Decreasing Redshift, $z$ \label{tab:radio-census}}
\tablewidth{0pt}
\tablehead{
\colhead{Name} & \colhead{$z$} & \colhead{Type} & \colhead{$m_{1450}$\tablenotemark{a}}  & \colhead{$\alpha_{\lambda,\mathrm{UV}}$\tablenotemark{b}}& \colhead{$f_{1.4\,\mathrm{GHz}}$} & \colhead{$\alpha_{\nu,\mathrm{radio}}$} & \colhead{$R_{2500}$\tablenotemark{c}} & \colhead{References}\\
 & \colhead{} & & \colhead{(mag)} & \colhead{} & \colhead{(mJy)} & & & \colhead{disc./$z$/$m_{1450}$/$\alpha_{\lambda,\mathrm{UV}}$/$f_{1.4\,\mathrm{GHz}}$/$\alpha_{\nu,\mathrm{radio}}$}
}
\startdata
\radioqso\  & 6.823 &  quasar & 21.08 & $-1.52$ & $0.510 \pm 0.016$ & $-1.31$ & $91 \pm 9$ & 1/1/1/1/1/1 \\
\hline 
J1429+5447  & 6.183 &  quasar & 20.70 & $-1.22$  & $2.93 \pm 0.15$ & $-0.67$ & $161 \pm 17$ & 2/3/4/--/5/6 \\
J1427+3312  & 6.121 &  quasar & 20.68 & --  & $1.73 \pm 0.13$ & $-0.90$ & $117 \pm 14$ & 7,8/9/4/--/5/6 \\
J0309+27172  & 6.10 &  blazar & 20.96 & --  & $23.89 \pm 0.87$ & $-0.44$ & $1521 \pm 151$ & 10/10/11/--/12/10 \\
J2228+0110  & 5.95  &  quasar & 22.20 & --  & $0.31 \pm 0.06$ & -- & $71 \pm 15$ & 13/13/13/--/13/-- \\
J2242+0334  & 5.88  &  quasar & 22.20 & --  & $0.20 \pm 0.03$ & $-1.06$ & $58 \pm 9$ & 2/2/4/--/14/14 \\
P352--15    & 5.84  &  quasar & 21.05 & --  & $14.9 \pm 0.70$ & $-0.89$ & $1358 \pm 141$ & 15/15/15/--/12/15 \\
J0836+0054  & 5.81  &  quasar & 18.95 & $-0.73$   & $1.74 \pm 0.04$ & $-0.86$ & $16 \pm 1$ & 16/17/4/9/18/6 \\
J1530+1049  & 5.72  &  radio galaxy & --    & --  & $7.50 \pm 0.10$ & $-1.40$ & -- & 19/19/--/--/19/19 \\
P055--00  & 5.68  &  quasar & 20.29    & --  & $2.14 \pm 0.14$ & -- & $83\pm 9$ & 20/20/4/--/5/-- \\
P135+16  & 5.63  &  quasar & 20.74    & --  & $3.04 \pm 0.15$ & -- & $177\pm 18$ & 20/20/4/--/5/--\\
J0856+0223  & 5.55  &  radio galaxy & --    & --  & $86.50 \pm 0.60$ & $-0.89$ & -- & 21/21/--/--/21/21 \\
J0906+6930  & 5.48 &  blazar & 19.67 & $-2.00$  & $92.0 \pm 0.62$ & $-0.40$ & $2373 \pm 205$ & 22/23/11/23/22/22 \\
J1648+4603  & 5.36 &  blazar & 19.51 & --  & $34.0 \pm 0.01$ & $-0.47$ & $552 \pm 47$ & 24/24/11/--/24/24 \\
J1614+4650  & 5.31 &  quasar & 19.72 & --  & $1.69 \pm 0.16$ & $0.67$ & $17 \pm 2$ & 24/25/11/--/5/6 \\
J1026+2542  & 5.25 &  blazar & 19.69 & --  & $230.00 \pm 0.14$ & $-0.60$ & $4701 \pm 407$ & 24/25/11/--/5/26\\
J2329+3003  & 5.24 &  quasar & 18.83 & --  & $4.90 \pm 0.40$ & -- & $47 \pm 5$ & 27/28/27/--/12/-- \\
J0924--2201  & 5.19  &  radio galaxy & --    & --  & $71.10 \pm 0.10$ & $-1.63$ & -- & 29/29/--/--/29/29  \\
J0131--0321  & 5.189 &  blazar & 18.09 & $-1.75$  & $32.83 \pm 0.12$ & $0.29$ & $116 \pm 9$ & 30/30/30/30/5/6 \\
J2245+0024  & 5.16 &  quasar & 22.24 & --  & $1.09 \pm 0.06$ & -- & $240 \pm 27$ & 31/31/31/--/32/-- \\
J0913+5919  & 5.12 &  quasar & 20.26 & --  & $17.45 \pm 0.16$ & $-0.67$ & $618 \pm 55$ & 24/25/11/--/5/33\\
J2239+0030  & 5.09 &  quasar & 21.27 & --  & $1.35 \pm 0.10$ & $-0.27$ & $98 \pm 12$ & 31/31/31/--/5/6 \\
J1034+2033  & 5.01 &  quasar & 19.56 & --  & $3.96 \pm 0.15$ & $0.28$ & $47 \pm 4$ & 24/25/11/--/5/6 \\
\hline
\hline
\enddata
\tablenotetext{a}{\footnotesize For objects for which the rest-frame 1450\,\AA\ magnitudes are not reported in the literature or have large uncertainties, we use as proxy their \yps\ magnitude from Pan-STARRS1 (Reference 11).} 
\tablenotetext{b}{\footnotesize We report rest-frame UV power-law slopes for objects with available near-infrared spectra covering at least from 1\,$\mu$m to 2.2\,$\mu$m. For J0836+0054, J1429+5447, and J0131--0321, $\alpha_{\lambda,\mathrm{UV}}$ was not directly available from the literature but we calculated it from their published spectra. 
} 
\tablenotetext{c}{\footnotesize To estimate $R_{2500}=f_{\nu,5\,\mathrm{GHz}} / f_{\nu,2500\,\text{\AA}}$, we extrapolate  $m_{1450}$  and  $f_{1.4\,\mathrm{GHz}}$ to rest-frame 2500\,\AA\ and 5\,GHz flux densities using the reported UV and radio slopes,  respectively. For objects without $\alpha_{\lambda,\mathrm{UV}}$, we assume the median value,$\alpha_{\lambda,\mathrm{UV,median}}=-1.36$ , found in the analysis of 38 $z\gtrsim 6$ quasars by \cite{schindler2020}. For objects without $\alpha_{\nu,\mathrm{radio}}$, we assume the median value  from all the `type=quasar' sources in this table: $\alpha_{\nu,\mathrm{radio,median}}=-0.67$.
See section \ref{sec:radio-loudness} for implications of extrapolating $\alpha_{\nu,\mathrm{radio}}$.
} 
\tablecomments{Blazars are highly variable objects and the UV and radio properties for the objects in this list were not observed simultaneously. Therefore, the radio-loudness reported here should be treated with caution, especially for blazars.}
\tablerefs{ 1: This work; 2: \cite{willott2010a}; 3: \cite{wang2011}; 
4: \cite{banados2016}; 5: \cite{becker1995}; 6: \cite{shao2020}; 
7: \cite{mcgreer2006}; 8: \cite{stern2007}; 9: \cite{shen2019};
10: \cite{belladitta2020}; 11: \yps\ magnitude; 12: \cite{condon1998}; 
13: \cite{zeimann2011}; 14: \cite{liu2021}; 
15: \cite{banados2018c};  16: \cite{fan2001};
17: \cite{kurk2007}; 18: \cite{wang2007}; 19: \cite{saxena2018};
20: \cite{banados2015a}; 21: \cite{drouart2020}; 
22: \cite{romani2004}; 23: \cite{an2018}; 
24: \cite{schneider2010}; 25: \cite{paris2018};
26: \cite{frey2015}; 27: \cite{wang-feige2016};
28: \cite{yang2016}; 29:  \cite{vanbreugel1999};
30: \cite{yi2014}; 31: \cite{mcgreer2013};
32: \cite{hodge2011}; 33: \cite{wu_j2013}
}
\end{deluxetable*}

\acknowledgments

E.B.\ thanks Aaron Meisner and Eddie Schlafly for helpful discussions about DECaLS and unWISE data. E.B.\ thanks Sof\'ia Rojas and Yana Khusanova for useful feedback on a previous version of this manuscript.  E.B.\ is extremely grateful to all the staff at Las Campanas Observatory for making all observing runs an unforgettable experience. 

C.M.\ acknowledges Eleonora Sani for useful discussion on quasar spectral fitting.

Research by A.J.B.\ is supported by NSF grant AST-1907290. 

A.C.E.\ acknowledges support by NASA through the NASA Hubble Fellowship grant $\#$HF2-51434 awarded by the Space Telescope Science Institute, which is operated by the Association of Universities for Research in Astronomy, Inc., for NASA, under contract NAS5-26555.

The work of T.C.\ and D.S.\ was carried out at the Jet Propulsion Laboratory, California Institute of Technology, under a contract with NASA. T.C.'s research was supported by an appointment to the NASA Postdoctoral Program at the Jet Propulsion Laboratory, California Institute of Technology, administered by Universities Space Research Association under contract with NASA.

B.P.V.\ acknowledges funding through the ERC Advanced Grant 740246 (Cosmic Gas).

We are grateful to the ESO and VLA staff for providing DDT observations for this program. 

The National Radio Astronomy Observatory is a facility of the National Science Foundation operated under cooperative agreement by Associated Universities, Inc.

This work is based on observations collected at the European Southern Observatory under ESO 
program 2102.A-5042(A). 
This paper includes data gathered with the 6.5 m Magellan Telescopes located at Las Campanas Observatory, Chile.

This paper includes data from the LBT. The LBT is an international collaboration among institutions in the United States, Italy, and Germany. The LBT Corporation partners are: The University of Arizona on behalf of the Arizona university system; Istituto Nazionale di Astrofisica, Italy;  LBT Beteiligungsgesellschaft, Germany, representing the Max Planck Society, the Astrophysical Institute Potsdam, and Heidelberg University; The Ohio State University; The Research Corporation, on behalf of The University of Notre Dame, University of Minnesota and University of Virginia.

Some of the data presented in this paper were obtained
at the W.M. Keck Observatory, which is operated as a scientific partnership among the California Institute
of Technology, the University of California and the National Aeronautics and Space Administration. The Observatory was made possible by the generous financial support of the W.M. Keck Foundation.  Keck data presented herein were obtained using the UCI Remote Observing Facility, made possible by a generous gift from John and Ruth Ann Evans. 

The authors wish to recognize and acknowledge the
very significant cultural role and reverence that the summit of Maunakea has always had within the indigenous Hawaiian community. We are most fortunate to have the opportunity to conduct observations from this mountain.

Based on observations made with the Nordic Optical Telescope, operated by the Nordic Optical Telescope Scientific Association at the Observatorio del Roque de los Muchachos, La Palma, Spain, of the Instituto de Astrofisica de Canarias.

\vspace{5mm}
\facilities{EVLA, Keck:II (NIRES), VLT:Kueyen (X-Shooter), Magellan:Baade (FIRE), LBT (MODS), NOT (NOTCam).}

\software{
Astropy \citep{astropy2018},
CASA \citep{mcmullin2007},
Matplotlib \citep[][\url{http://www.matplotlib.org}]{hunter2007},
Numpy \citep{harris2020},
PypeIt \citep{prochaska2019,prochaska2020},
SciPy \citep{virtanen2020}, 
TOPCAT \citep[][\url{http://www.starlink.ac.uk/topcat/}]{taylor2005}
          }

\bibliographystyle{aasjournal}

\end{document}